\documentclass[aps,prb,showpacs,superscriptaddress,twocolumn]{revtex4}%,longbibliography
\usepackage{amsfonts}
\usepackage{txfonts}
\usepackage{amssymb}
\usepackage{amsbsy} %for sigma in bold face
\usepackage{epsfig}
\usepackage{subfigure}

\def\be{\begin{equation}} \def\ee{\end{equation}}
\def\bea{\begin{eqnarray}} \def\eea{\end{eqnarray}}

\def\nn{\nonumber}

\def\bq{{\bf q}}
\def\bQ{{\bf Q}}
\def\bk{{\bf k}}

\def\bp{{\bf p}}

\def\bB{{\bf B}}

\def\bK{{\bf K}}
\def\be{{\bf e}}

\def\bL{{\bf L}}

\def\la{\langle}
\def\ra{\rangle}

\def\rw{\rightarrow}

\begin{document}

\title{Unidirectional Transport in Electronic and Photonic Weyl Materials by Dirac Mass Engineering}

\author{Ren Bi} \affiliation{ Institute for
Advanced Study, Tsinghua University, Beijing,  China, 100084}

\author{Zhong Wang}
\altaffiliation{ Corresponding author. wangzhongemail@tsinghua.edu.cn} \affiliation{ Institute for
Advanced Study, Tsinghua University, Beijing,  China, 100084}

\affiliation{Collaborative Innovation Center of Quantum Matter, Beijing 100871, China }

\date{ \today}

\begin{abstract}

Unidirectional transports have been observed in two-dimensional systems, however, so far they have not been experimentally observed in three-dimensional bulk materials. In this theoretical work we show that the recently discovered Weyl materials provide a platform for unidirectional transports inside bulk materials. With high experimental feasibility, a complex Dirac mass can be generated and manipulated in the photonic Weyl crystals, creating unidirectionally propagating modes observable in transmission experiments. Possible realization in (electronic) Weyl semimetals is also studied. We show in a lattice model that, with a short-range interaction, the desired form of the Dirac mass can be spontaneously generated in a first-order transition.

\end{abstract}

\pacs{73.43.-f,78.67.Pt}

\maketitle

\emph{Introduction.}
Recently, novel materials described by the Weyl equation are among the research focuses in physics\cite{wan2011,NIELSEN1981,nielsen1983adler,volovik2003,burkov2011}. For instance, the phenomenon of chiral anomaly\cite{Adler-anomaly,bell1969} has been extensively investigated in Weyl semimetals \cite{son2012,liu2012,aji2011,wang2013a,zyuzin2012,Hosur-anomaly,
Hosur2013,Kim-chiral-anomaly,Parameswaran-anomaly, Zhou-plasmon,Li2015ZrTe5,Goswami2015}.
The recent experimental discoveries of the photonic Weyl crystals\cite{lu2015,lu2013weyl,Lu-review} and the TaAs
class Weyl semimetals\cite{weng2015,Huang2015TaAs,Zhang2015a,xu2015,lv2015,Huang2015,
YangLexian,Ghimire,Shekhar,Xu2015NbAs}
have generated intense interests in Weyl materials.

\begin{figure}
\centering
\includegraphics[width=8cm, height=5cm]{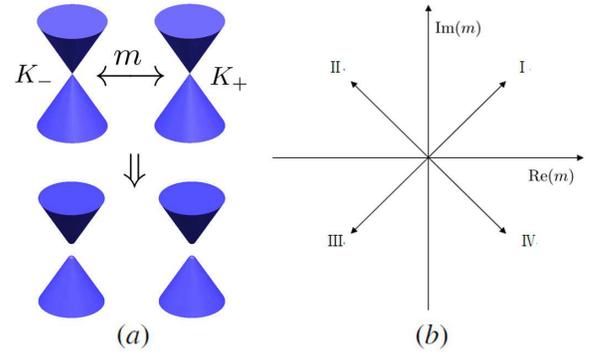}
\caption{  (a) Coupling two Weyl points by a Dirac mass term. (b) A ``mass clock'' showing the four values of the complex Dirac mass in our lattice model. } \label{clock}
\end{figure}

A Dirac mass can couple
two Weyl points together to form a massive Dirac fermion (Fig.\ref{clock}a)\footnote{This can be achieved by spontaneous mass generation \cite{fradkin1983,zyuzin2012,wang2013a,wei2012,maciejko2014,roy2014, Sun2015helical}.  }.
The
Dirac mass is generally complex-valued, and it is known in continuous field theory that the vortex lines of mass carry unidirectionally propagating modes\cite{jackiw1981zero,weinberg1981index,witten1985superconducting, callan1985anomalies,Harvey2005}. Since the unidirectional modes are completely immune to backscattering, they are capable of supporting dissipationless transport. Such unidirectional modes were found at the edge of quantum Hall systems and their photonic analogues\cite{Haldane-optical,Raghu-photonic,wang2009observation}, however, so far they have not been experimentally observed inside three-dimensional (3D) bulk materials\footnote{For conceptual discussions, see Ref.\cite{teo2010,wang2013a}}. Therefore, it is highly desirable to seek experimental realization of the vortex line of Dirac mass, which is nevertheless challenging. In particular, designing a material with the phase of mass continuously tunable from $0$ to $2\pi$ is a difficult task\footnote{Vortex line of Dirac mass in the Weyl semimetal has been briefly discussed in Ref.\cite{wang2013a}, though no concrete model was studied there. }.

The results of this paper are threefold. First, we study the unidirectional modes in 3D \emph{lattice} Weyl materials, and construct a discrete version of the vortex line of mass. This study is complementary to the continuous theories mentioned previously. New features absent in the continuous theories are analyzed.

Second, we propose to realize the vortex line of mass and the associated unidirectional modes in \emph{photonic} Weyl crystals. This can be achieved by a simple modulation of the photonic Weyl crystals, which naturally generates a complex Dirac mass, whose phase can be manipulated by lattice translations. This proposal is most feasible after the recent experimental realization\cite{lu2015} of the gyroid photonic Weyl crystals. This is our central proposal.

Third, we point out that the spontaneous mass generation in Weyl semimetals can be first-order due to the lattice effect. This first-order nature is absent in continuous approximations with a momentum cutoff (e.g.  Ref.\cite{wei2012}).

\begin{figure}
\centering
\includegraphics[width=8cm, height=6cm]{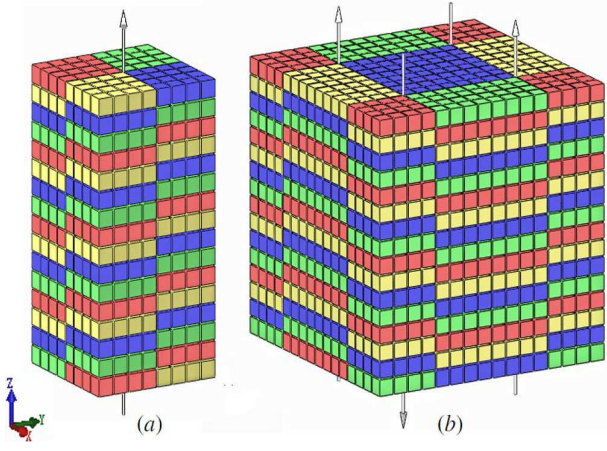}
\caption{ Lattice system with dislocation lines.  Each small cube represents a unit cell. Different colors represent different values of $m_i$ (namely, $m_1, m_2, m_3, m_4$, see text).  (a) A vortex line of the complex Dirac mass (axion string). (b) A system with four vortex lines.  Periodic boundary condition is taken in the $x$ and $y$ direction, while $k_z$ is kept as a good quantum number. The arrows indicate the propagating directions of the unidirectional modes. } \label{dislocation}
\end{figure}

\emph{Lattice model and Dirac mass engineering.} Let us begin with a simplest lattice model, which captures the key low-energy physics of Weyl semimetals and photonic Weyl materials with a single pair of Weyl points. The lattice
Hamiltonian\cite{yang2011} is given as \bea  H_0(\bk)  = && 2t_x\sin
k_x\sigma_x + 2t_y\sin k_y\sigma_y + [2t_z(\cos k_z-\cos k_0)  \nn \\
&& + B(2-\cos k_x -\cos k_y)]\sigma_z. \label{lattice} \eea For
simplicity, the lattice constants $a$ has been
scaled to $1$, therefore $k_i$ has periodicity $2\pi/a=2\pi$. We take the interpretation that each unit cell
contains a single site, and the Pauli matrix $\sigma_i$ refers to the spin  (or pseudospin) degree of
freedom. This
lattice model possesses two Weyl points at $\bK_\pm=(0,0,\mp k_0)$,
with difference $\bQ\equiv \bK_- -\bK_+ = (0, 0, Q) = (0, 0,  2k_0)$.
Near the Weyl points, we have the low-energy Hamiltonian \bea H_W
(\bq) =v_x\sigma_x q_x + v_y\sigma_y q_y + v_z\sigma_z\tau_z q_z,
\label{hW} \eea where $v_{x(y)}=2t_{x(y)}$, $v_z=2t_z\sin k_0$, the
Pauli matrix $\tau_z$ is the chirality operator, namely, the
eigenvalue of $\tau_z$ is $\pm 1$ for Bloch states near $\bK_\pm$, and
$\bk-\bK_\pm\equiv (q_x, q_y, q_z)$ is the momentum measured relative
to the Weyl points. Hereafter we shall take $k_0=\pi/4$,
$t_x=t_y=0.5$, $t_z=1$, $B=2.5$, thus $\bQ =(0, 0, \pi/2)$.

The Dirac mass terms can be formally included as \bea H_M = m\sigma_z\tau_+  + m^*\sigma_z\tau_- ,\eea where $\tau_\pm \equiv (\tau_x\pm i\tau_y)/2$, and both $\sigma_z\tau_+$ and $\sigma_z\tau_-$ anticommute with the three terms in $H_W$. At low energy, the eigenvalues of the full Hamiltonian $H_W+H_M$ are $\pm\sqrt{\sum_{i=x,y,z} v_i^2 q_i^2+|m|^2}$, with a gap $2|m|$.

Now we would like to introduce a Dirac mass in the \emph{lattice} model in Eq.(\ref{lattice}). Because the momentum difference between the two Weyl points is $\bQ=(0,0,\pi/2)$, a Dirac mass must break the translational symmetry, enlarging the size of unit cell by a factor of four in the $z$ direction. The four adjacent sites in each enlarged unit cell are sequentially denoted as $A, B, C, D$, namely, ``$A$'' for $z=4n$, ``$B$'' for $z=4n+1$, ``$C$'' for $z=4n+2$, and ``$D$'' for $z=4n+3$ ($n$ being integers), where $z$ refers to the $z$ coordinate. To generate a Dirac mass, we add a translational-symmetry-breaking term into the Hamiltonian: \bea H'=\sum_j m_j\sigma_z(j), \label{lattice-mass} \eea where $j$ runs through all sites, and $m_j = m_A, m_B, m_C, m_D$ for $A, B, C, D$ site, respectively. The inclusion of $\sigma_z$ in $H'$ is suggested by the form of $H_M$.

After enlargement of the unit cell, the lattice Hamiltonians $H_0$ and $H'$ are replaced by eight-band lattice Hamiltonians $\bar{H}_0$ and $\bar{H}'$ (Supplemental Material),  meanwhile, the Brillouin zone of the enlarged unit cell is $[0,2\pi]\times[0,2\pi]\times[0,\pi/2]$. We find that the energy eigenvalues of $\bar{H}_0$ have fourfold degeneracies at zero energy at $\bk=(0,0,\pi/4)$. For these four low-energy bands we obtain (Supplemental Material) the following effective Hamiltonian of $\bar{H}_0+\bar{H}'$:
\bea \bar{H}_{\rm eff}
(\bq) = && v_x\sigma_x q_x + v_y\sigma_y q_y + v_z\sigma_z\tau_z  q_z   + m\sigma_z\tau_+  + m^*\sigma_z\tau_- \nn \\ && +\Delta\sigma_z,  \label{Heff} \eea where $(q_x,q_y,q_z)\equiv (k_x,k_y,k_z)-(0,0,\pi/4)$, $v_{x(y)}=2t_{x(y)}$, $v_z=2t_z\sin(\pi/4)$, $\Delta=(m_A+m_B+m_C+m_D)/4$, and \bea m=(m_A+e^{iQ}m_B + e^{2iQ}m_C +e^{3iQ}m_D)/4. \eea
We shall study the $\Delta$ term later; at this stage we just omit it and focus on the Dirac mass $m$.

Now we can construct a vortex line of the Dirac mass. We observe that for the four cases, (I) $(m_A, m_B, m_C, m_D)=(m_1,m_2,m_3,m_4)$; (II) $(m_A, m_B, m_C, m_D)=(m_4,m_1,m_2,m_3 )$; (III) $(m_A, m_B, m_C, m_D)=(m_3,m_4,m_1,m_2)$; (IV) $(m_A, m_B, m_C, m_D)=( m_2,m_3,m_4,m_1)$, the Dirac mass is given by $m_{\rm I}=(m_1+im_2   -m_3 -i m_4)/4$, $m_{\rm II}= (m_4+im_1   -m_2 -i m_3)/4=im_{\rm I}$, $m_{\rm III}= (m_3+im_4   -m_1 -i m_2)/4=-m_{\rm I}$, and $m_{\rm IV}= (m_2+im_3   -m_4 -i m_1)/4=-im_{\rm I}$, respectively.
We can see that $m$ is multiplied by a factor $i$
during each move in I$\rw$II$\rw$III$\rw$IV$\rw$I. It is worth noting
that configurations I, II, III, and IV are related by translations
along the $z$ direction\footnote{In the Dirac semimetals\cite{liu2014discovery,neupane2014,Borisenko2014,xu2015observation, wang2012dirac,young2012dirac,wang2013three,Sekine2014,Zhang2015detection}, there is no such a relation between the Dirac mass and lattice translations.}. There is no difference in physical
properties to distinguish them, unless they are adjacent to each other. Nevertheless, we can design a
vortex line (an ``axion string''), such that along a circle enclosing
this dislocation, the path I$\rw$II$\rw$III$\rw$IV$\rw$I is followed
(see Fig.\ref{dislocation}a ). This is a discrete version of the vortex line of Dirac mass (see Fig.\ref{clock}b).

Hereafter we shall take $(m_1,m_2,m_3,m_4)=(m_0,m_0,-m_0,-m_0)$, which leads to $\Delta=0$. The Dirac mass for the four configurations is $m_{\rm I}= \frac{m_0}{\sqrt{2}} \exp(i\pi/4)$, $m_{\rm II}=im_{\rm I}$, $m_{\rm III}=- m_{\rm I}$, and $m_{\rm IV}=-im_{\rm I}$, respectively.

\begin{figure}
\centering
\includegraphics[width=8cm, height=4.5cm]{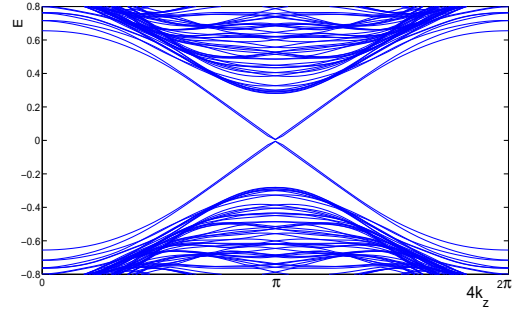}
\caption{ The spectrum of the system illustrated in Fig.\ref{dislocation}b. The system size is $28\times 28$ unit cell in the $xy$ plane, and $m_0=0.5$. Four dislocation modes inside the bulk energy gap can be clearly seen. (The tiny energy gap at $4k_z=\pi$ is due to finite-size effect.)
Energy bands far from zero-energy are not shown here. } \label{photonic-spectrum}
\end{figure}

In the continuous field
theory (for instance, see Ref.\cite{callan1985anomalies}), the vortex line of $m$ carries a unidirectional
mode. It is conceivable that our discrete version of vortex line of $m$ also carries unidirectional mode.
To be conclusive, we do a numerical calculation on a system with periodic boundary condition in the $x$ and $y$ directions (see Fig.\ref{dislocation}b). In the calculation we have Fourier transformed $H_0$ in Eq.(\ref{lattice}) into real space in the $x$ and $y$ directions.  The energy spectrum (of the full Hamiltonian $\bar{H}\equiv \bar{H}_0+\bar{H}'$) as a function of $k_z$ is plotted in Fig.\ref{photonic-spectrum}, from which we can see four unidirectional modes, two $+z$ propagating and two $-z$ propagating.

\begin{figure}
\subfigure{\includegraphics[width=8cm, height=4.5cm]{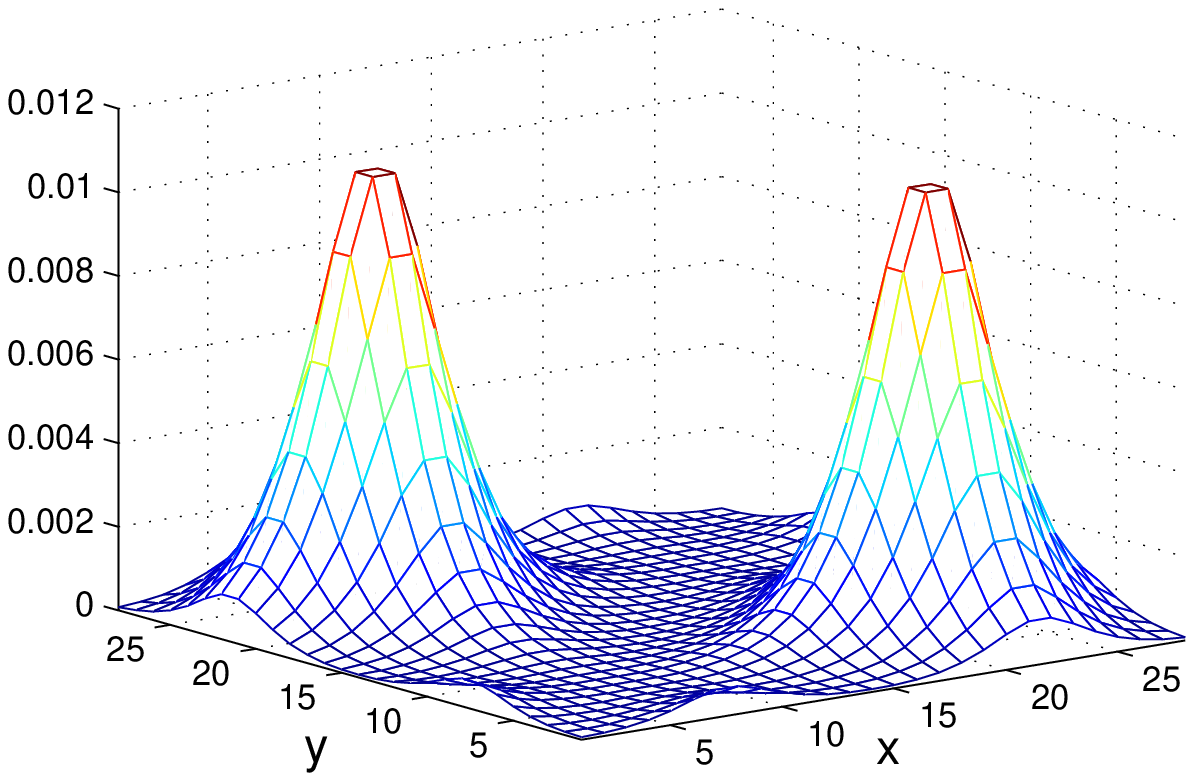}}
\subfigure{\includegraphics[width=8cm, height=4.5cm]{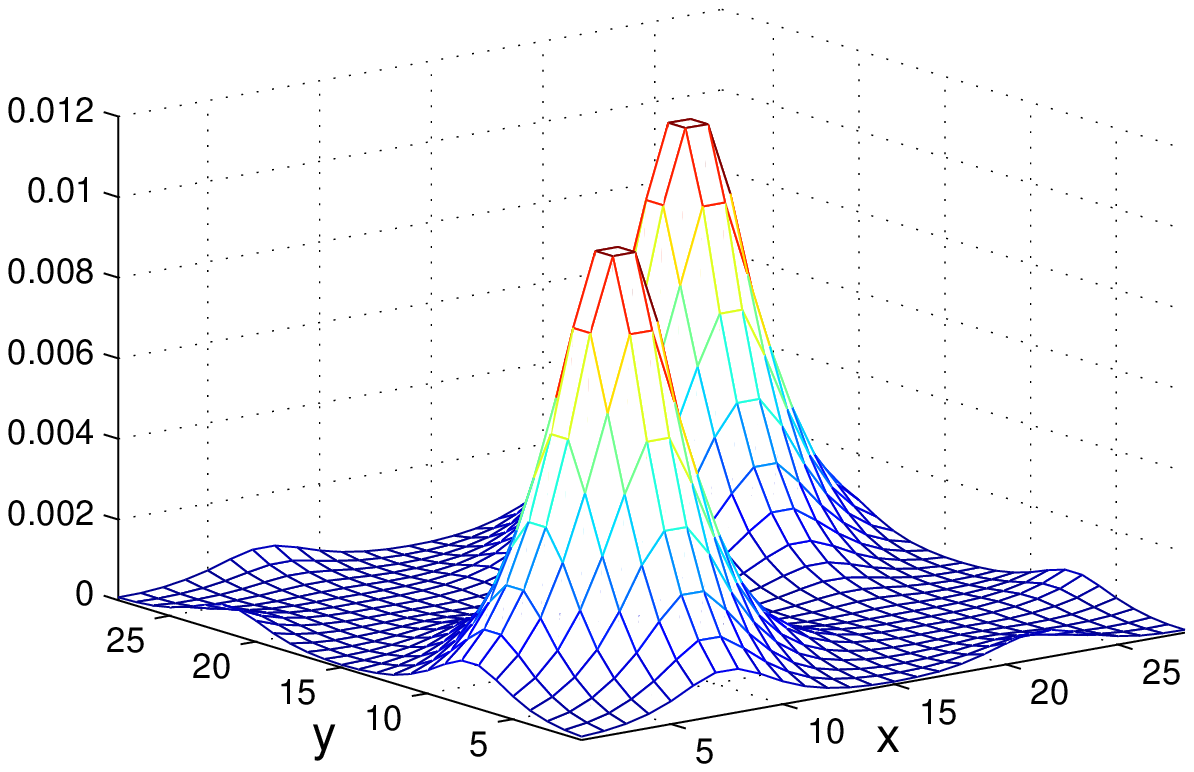}}
\caption{   The density distributions of the unidirectional modes propagating in the $+z$ (Upper) and $-z$ (Lower) directions, at momentum $k_z=0.97\pi/4$. The parameters are the same as specified in Fig.\ref{photonic-spectrum}.  }\label{density}
\end{figure}

To show that the unidirectional modes are indeed localized around the dislocation lines, we plot their density distributions in Fig.\ref{density}. We only plot one mode for each direction of propagation ($+z$ and $-z$). Each mode is evenly distributed around two dislocation lines because of the tunneling effect (like in the double-well quantum mechanical system).

Previously we have omitted the $\Delta$ term in Eq.(\ref{Heff}). With this term included, the energy eigenvalues are \bea E_{1,2,3,4}(\bq)= \pm \sqrt{v_x^2 q_x^2 +v_y^2 q_y^2 + (\Delta \pm\sqrt{v_z^2 q_z^2+|m|^2})^2},\eea which become gapless when $|\Delta|\geq |m| $. As long as $|\Delta|<|m| $, the bulk material is gapped, and the topologically robust unidirectional modes persist.

What if $2\pi/|\bQ|$ is incommensurate with the lattice constant?  Let us take $k_0=\pi/4+\delta$, with $\delta$ being small. This changes $\Delta$ in Eq.(\ref{Heff}) to $\Delta  -2t_z\cos (\pi/4+\delta) + 2t_z\cos(\pi/4)\approx \Delta + v_z\delta$ (Supplemental Material). The previous condition $|\Delta|<|m| $ is then replaced by $|\Delta + v_z\delta|< |m| $. Therefore, a small $\delta$ does not qualitatively change our result.

The above study is relevant to Weyl semimetals with time-reversal
symmetry breaking, such as magnetic Weyl
semimetals\cite{CavaWeyl,xu2011}.  In the next
section we will study electron-electron interaction that may
spontaneously generate the desired form of mass. More feasibly, we propose to realize the
unidirectional modes in photonic Weyl materials, for
which the Dirac mass can be readily generated and controlled.  In the gyroid
photonic crystals studied in Ref.\cite{lu2015}(Supplemental Material), one can tune $\bQ$ to
be close to $(0, 0, \pi/2)$, and produce a modulation with a period of
four unit cells (by making the $A, B$ unit cells slightly different
from the $C, D$ unit cells). From symmetry consideration, this will generate the $H'$ term at low energy, which leads to a Dirac mass. Then one can fabricate vortex lines (see Fig.\ref{dislocation}) and observe unidirectional propagation of
light (or microwave) in transmission experiments. The discrete version of vortex line studied here is convenient to fabricate in experiments.

\emph{Spontaneously generated mass.} In this section we study an interacting electron model, whose free Hamiltonian is $\hat{H}_0=\sum_\bk c^\dag_\bk H_0(\bk) c_\bk$, where $c_\bk$ is the electron operator, and
\bea H_0(\bk) &=& 2t_x\sin   k_x\sigma_x +2t_y\sin k_y\sigma_y   +
 [2t_z(\cos k_z - \cos k_0)  \nn \\ && + B(\cos k_x -\cos k_y)^2]
   \sigma_z. \label{lattice-2} \eea
We shall take $t_x=t_y=0.2, t_z=1, B=1.48$, and
$k_0=\pi/4$. The Brillouin zone contains four Weyl points,
$\bK_{\pm}=(0,0,\mp k_0)$ and $\bK'_{\pm}=(\pi,\pi,\mp k_0)$.  The
low energy electrons near $\bK_\pm$ can be described by Eq.(\ref{hW}), while those
near $K'_\pm$ can be described by \bea H'_W
(\bq) = -v_x\sigma_x q_x - v_y\sigma_y q_y + v_z\sigma_z\tau_z q_z,
\label{hW'} \eea In the continuous theory, it is well known that spontaneous mass generation can be induced by a four-fermion interaction\cite{nambu1961}, which suggests us to consider
a Hubbard-like interaction $\hat{H}_I=-U\sum_j (c^\dag_j\sigma_z c_j)^2$ in our model. It can be decoupled as $\sum_j(- 2U M_j  c^\dag_j\sigma_z c_j + U M_j^2)$ at the mean-field level\cite{fradkin2013}, where we have introduced
$M_j \equiv \la c^\dag_j \sigma_z c_j\ra$. The mean-field Hamiltonian is therefore \bea \hat{H}_{{\rm MF}} = \hat{H}_0 - 2U \sum_j   M_j  c^\dag_j\sigma_z c_j , \label{MF} \eea Our mean-field calculation shall be quite standard, except that we shall enlarge the unit cell to $L_x\times L_y\times L_z$  to accommodate possible symmetry-breaking phases ( $L_x\times L_y\times L_z=1\times 1\times 4$, $2\times 2\times 4$, $1\times 1\times 8$, and $2\times 2\times 8$ are calculated). Since $\hat{H}_{{\rm MF}}$ is quadratic, it can be diagonalized in the $\bk$ space, with eigenvalues $\epsilon_\alpha(\bk)$, and the mean-field energy per unit cell is $E_{\rm MF}=\int\frac{d^3k}{(2\pi)^3}\sum_{\alpha}\epsilon_\alpha(\bk) + U\sum_{i=1}^{L_xL_yL_z} M_i^2$, where the summation $\sum_\alpha$ is over all negative energies $\epsilon_\alpha(\bk)<0$, i.e. the energies of the occupied Bloch states. The mean-field ground-state can be obtained by minimizing $E_{\rm MF}$ with respect to $M_i$, leading to standard integral equations \footnote{Analogous to Eq.(3.45) of Ref.\cite{fradkin2013}.} for $M_i$ ($i=1,2,\cdots,L_xL_yL_z$), which can be solved iteratively in numerical calculations\footnote{In the cases of several solutions, we should take the one with lowest mean-field energy as the ground-state.}.

\begin{figure}
\centering
\includegraphics[width=7cm, height=5cm]{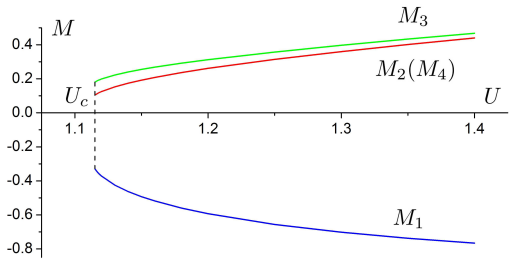}
\caption{ The ground-state $M_i$'s as functions of the interaction parameter $U$. The Dirac mass is nonzero when $U>U_c$ (see text). } \label{UM}
\end{figure}

For the $1\times 1\times 4$ unit cell, the obtained values of $(M_1,M_2,M_3,M_4)$ are plotted in Fig.\ref{UM}. All $M_i$'s are zero for small $U$. At $U_c\approx 1.12$, there is a first-order transition to a symmetry-breaking phase
with $M_i\neq 0$. The same phase is reproduced in calculations on larger unit cells. For instance, we find that the values of $\{M_i\}$ are $(M_1,M_2,M_3,M_4,M_1,M_2,M_3,M_4)$ for the $1\times 1\times 8$ unit cell, with the same values of $M_{1,2,3,4}$ as obtained from the $1\times 1\times 4$ unit cell.
A first-order transition is understandable here because $E_{\rm MF}(M_i)\neq E_{\rm MF}(-M_i)$, so that $E_{\rm MF}$ contains odd-order terms when expanded in powers of $M_i$'s (similar to the Ginzburg-Landau theory of first-order nematic-to-isotropic transition in liquid crystals).

\begin{figure}
\subfigure{\includegraphics[width=8cm, height=4.5cm]{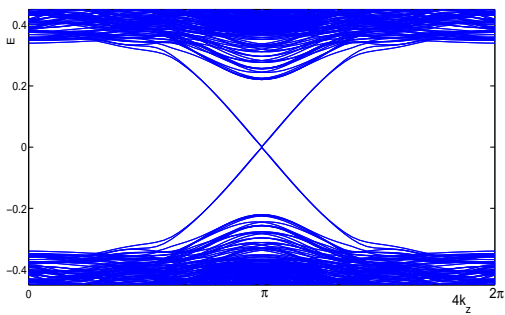}}
\subfigure{\includegraphics[width=8cm, height=4.5cm]{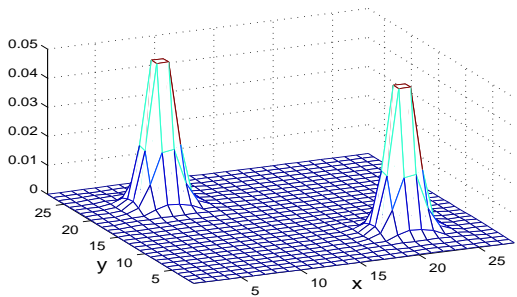}}
\caption{ Upper: The spectrum of the system with dislocation lines illustrated in Fig.\ref{dislocation}b, for the mean-field Hamiltonian $\hat{H}_{{\rm MF}}$ at $U=1.125$. The system size is $28\times 28$ in the $xy$ plane. Lower: The density distribution of one of the unidirectional modes (at $k_z=0.97\pi/4$). } \label{spectrum-two-Weyl}
\end{figure}

Comparing Eq.(\ref{MF}) with Eq.(\ref{lattice-mass}), we can see that the physics of the previous section applies here if we identify $m_j=-2UM_j$; in particular, the Dirac mass is $m=-\frac{U}{2}(M_A+iM_B-M_C-iM_D)$. Dislocation lines can be constructed in similar way to the previous section.  The spectrum and the density distribution of one of the eight unidirectional modes is shown in Fig.\ref{spectrum-two-Weyl} (Due to the presence of \emph{two} copies of Dirac fermions, the number of unidirectional modes is twice that of the previous model).

Finally, we remark that dynamical mass generation is not the only approach to create a Dirac mass. In the Weyl semimetals based on layered structures\cite{burkov2011}, we suggest to introduce an additional superlattice structure with period near $2\pi/|\bQ|$ (say a modulation of the magnetization with period near $2\pi/|\bQ|$), which will generate a Dirac mass.  Then vortex lines of the Dirac mass can be constructed (similar to Fig.\ref{dislocation}).

\emph{Final remarks.}
We have shown how to create unidirectional modes in the bulk materials of electronic and photonic Weyl materials.
Compared to the helical dislocation modes in weak topological insulators\cite{Ran2008,Imura2011,Slager2014}, our unidirectional modes are completely immune to any kind of backscattering. In this respect our proposal will be of significant practical use.

We have proposed the photonic Weyl materials as the most experimentally feasible systems, because the Dirac mass can be readily designed and controlled therein. The recent experimental observation of photonic Weyl points\cite{lu2015} is most propitious to this proposal. Finally, we remark that our proposal may be generalized to other topological materials (e.g. topological phononic or acoustic materials\cite{Susstrunk2015,Prodan2009phonon,kane2014topological, Peano1015topological,Yang2015acoustics,paulose2015topological,Weyl-acoustic,Wang2015phononic}).

\emph{Acknowledgements.}  It is our pleasure to thank Ling Lu for
helpful discussions about the experimental feasibility. This work is
supported by NSFC under Grant No. 11304175 and Tsinghua University
Initiative Scientific Research Program.

\bibliography{dirac}

\vspace{8mm}

{\bf Supplemental Material}

\section{Dirac Hamiltonian and Dirac mass in the enlarged unit cell}\label{calculation}

In the paper, we have seen that the translational-symmetry-breaking term $H'=\sum_j m_j\sigma_z(j)$ enlarges the size of the unit cell by a factor of four in the $z$ direction [Note: $\sigma_z(j)$ is the Pauli matrix $\sigma_z$ defined at the $j$ site]. In the enlarged unit cell, we have $2\times 4=8$ bands. Now we derive the low-energy Dirac Hamiltonian and the Dirac mass.

For the enlarged unit cell, the first Brillouin zone is $[0,2\pi]\times[0,2\pi]\times[0,\pi/2]$ (the lattice constant $a$ of the original unit cell is taken to be $1$).
According to the standard band theory, within the description of the enlarged unit cell, the lattice Hamiltonian $H_0$ becomes  \bea \bar{H}_{0}(\bk) =  \left( \begin{array}{cccc}
        H_{11} & H_{12} & H_{13} & H_{14}\\
         H_{21} & H_{22} & H_{23} & H_{24}\\
         H_{31} & H_{32} & H_{33} & H_{34}\\
         H_{41} & H_{42} & H_{43} & H_{44}\\
         \end{array}
\right),
         \eea
         where $H_{ij}$'s are $2\times 2$ matrices: \bea && H_{11} = H_{22}=H_{33}=H_{44}  = 2t_x\sin k_x\sigma_x + 2t_y \sin k_y\sigma_y  \nn \\ && +[-2t_z\cos k_0 + B(2-\cos k_x -\cos k_y)]\sigma_z, \nn \\ &&
         H_{12}=H_{21}=H_{23}=H_{32}=H_{34}=H_{43}=t_z\sigma_z, \nn \\ &&
         H_{14}=H^\dag_{41}=t_z\exp(-4ik_z)\sigma_z,\nn \\
         && H_{13}=H_{31}=H_{24}=H_{42}=0. \eea
The eight components of the state vectors have been ordered as $(A\uparrow,A\downarrow,B\uparrow,B\downarrow,C\uparrow,C\downarrow,D\uparrow,D\downarrow)$, $\uparrow$ and $\downarrow$ referring to the two eigenstates of $\sigma_z$.
As specified in the paper, we take $k_0=\pi/4$,
$t_x=t_y=0.5$, $t_z=1$, $B=2.5$. The only gapless point in the Brillouin zone is $(0,0,\pi/4)$ [Note that $(0,0,-\pi/4)$ is the same point because $k_z$ has periodicity $\pi/2$]. We can find that $\bar{H}_0$ have four zero eigenvalues at $(0,0,\pi/4)$, the corresponding eigenvectors are \bea  |\psi_1\ra &=& \frac{1}{2}\left( \begin{array}{c}
        1 \\
        0 \\
        e^{-i\pi/4}\\
         0 \\
         e^{-i\pi/2} \\
         0 \\
         e^{-3i\pi/4}\\
         0 \\
         \end{array}
\right), \,\,
|\psi_2\ra = \frac{1}{2}\left( \begin{array}{c}
        0 \\
        1 \\
        0 \\
         e^{-i\pi/4}\\
         0 \\
         e^{-i\pi/2} \\
         0 \\
         e^{-3i\pi/4}\\
         \end{array}
\right), \nn \\
 |\psi_3\ra &=& \frac{1}{2}\left( \begin{array}{c}
        1 \\
        0 \\
        e^{i\pi/4}\\
         0 \\
         e^{i\pi/2} \\
         0 \\
         e^{3i\pi/4}\\
         0 \\
         \end{array}
\right), \,\,
|\psi_4\ra = \frac{1}{2}\left( \begin{array}{c}
        0 \\
        1 \\
        0 \\
         e^{i\pi/4}\\
         0 \\
         e^{i\pi/2} \\
         0 \\
         e^{3i\pi/4}\\
         \end{array}
\right). \eea
The existence of four-fold degeneracy at zero energy is understandable because the two Weyl points in the original small unit cell are folded to the same point in the Brillouin zone of the enlarged unit cell. Here, $|\psi_1\ra$ and $|\psi_2\ra$ have $\sigma_z=+1$ and $-1$, respectively, and similarly for $|\psi_3\ra$ and $|\psi_4\ra$. For later convenience, we also introduce a quantum number $\tau_z$, which is $+1$ for $|\psi_1\ra$ and $|\psi_2\ra$, $-1$ for $|\psi_3\ra$ and $|\psi_4\ra$. We shall not be concerned with the other four eigenvectors with energies of order of $t_z$.

Near the gapless point $(0,0,\pi/4)$, for the four low-energy bands,  we can use the standard $\bk\cdot\bp$ perturbation theory\cite{cardona2005fundamentals} to obtain a low-energy effective Hamiltonian $\bar{H}_{0,{\rm eff}}$, whose entries are \bea  (\bar{H}_{0,{\rm eff}}(\bq))_{\alpha\beta} = \sum_{i=x,y,z}q_i \la\psi_\alpha|\frac{\partial\bar{H}_0}{\partial k_i}|\psi_\beta\ra +\cdots \eea where $(q_x,q_y,q_z)\equiv (k_x,k_y,k_z)-(0,0,\pi/4)$,  $\alpha,\beta=1,2,3,4$,  the partial derivatives are taken at $(0,0,\pi/4)$, and ``$\cdots$'' refers to higher order terms, which are omitted in the low-energy linear approximation. These matrix elements can be calculated explicitly. For instance,   \bea \frac{\partial\bar{H}_0}{\partial k_z}= \left( \begin{array}{cccc}
        0 & 0 & 0 & 4it_z\sigma_z \\
         0 & 0 & 0 & 0\\
         0 & 0 & 0 & 0\\
         -4it_z\sigma_z & 0 & 0 & 0\\
         \end{array}
\right), \eea therefore, we have \bea \la\psi_1|\frac{\partial\bar{H}_0}{\partial k_z}|\psi_1\ra = - \la\psi_2|\frac{\partial\bar{H}_0}{\partial k_z}|\psi_2\ra = \la\psi_3|\frac{\partial\bar{H}_0}{\partial k_z}|\psi_3\ra \nn \\ = -\la\psi_4|\frac{\partial\bar{H}_0}{\partial k_z}|\psi_4\ra = 2t_z\sin(\pi/4), \eea while $\la\psi_\alpha|\frac{\partial\bar{H}_0}{\partial k_z}|\psi_\beta\ra=0$ for other $\alpha,\beta$ pairs. All other matrix elements can be calculated similarly, and the final result can be written concisely as
\bea \bar{H}_{0,{\rm eff}} = v_x q_x \sigma_x + v_y q_y \sigma_y + v_z q_z \sigma_z \tau_z, \eea where $v_{x(y)}=2t_{x(y)}$, $v_z=2t_z\sin(\pi/4)$. This is a massless Dirac Hamiltonian.

Next we add the $H'=\sum_j m_j\sigma_z(j)$ term. In the eight-band description of the enlarged unit cell, this mass term reads \bea \bar{H}'= \left( \begin{array}{cccc}
        m_A\sigma_z & 0 & 0 & 0 \\
         0 & m_B\sigma_z & 0 & 0\\
         0 & 0 & m_C \sigma_z & 0\\
         0 & 0 & 0 & m_D\sigma_z \\
         \end{array}
\right). \eea Near the gapless point of $\bar{H}_{0,{\rm eff}}$, namely,  $(0,0,\pi/4)$ ,  we would like to obtain its effective   form within the four low-energy bands, denoted by $\bar{H}'_{\rm eff}$, whose matrix entries are \bea (\bar{H}'_{\rm eff})_{\alpha\beta} = \la \psi_\alpha|\bar{H}'|\psi_\beta\ra. \eea By explicit calculations, it is not difficult to find that \bea \la\psi_1|\bar{H}'|\psi_1\ra =  -\la\psi_2|\bar{H}'|\psi_2\ra =  \la\psi_3|\bar{H}'|\psi_3\ra \nn \\ = -\la\psi_4|\bar{H}'|\psi_4\ra =\frac{1}{4}(m_A+m_B+m_C+m_D),  \eea and \bea  \la\psi_1|\bar{H}'|\psi_3\ra  && =\frac{1}{4}(m_A+e^{i\pi/2}m_B+ e^{i\pi}m_C+e^{3i\pi/2}m_D) \nn \\ && = \frac{1}{4}(m_A+i m_B-m_C -im_D), \nn \\
 \la\psi_2|\bar{H}'|\psi_4\ra  && = -\frac{1}{4}(m_A+e^{i\pi/2}m_B+ e^{i\pi}m_C+e^{3i\pi/2}m_D) \nn \\ && = -\frac{1}{4}(m_A+i m_B-m_C -im_D),  \nn \\  \la\psi_3|\bar{H}'|\psi_1\ra && =(\la\psi_1|\bar{H}'|\psi_3\ra)^*, \nn \\ \la\psi_4|\bar{H}'|\psi_2\ra && =(\la\psi_2|\bar{H}'|\psi_4\ra)^* . \eea All other matrix elements of $\la \psi_\alpha|\bar{H}'|\psi_\beta\ra$ vanish. Therefore, at low-energy the Dirac mass term $H'$ takes the form \bea \bar{H}'_{\rm eff} =  \left( \begin{array}{cccc}
        \Delta & 0 & m & 0 \\
         0 & -\Delta & 0 & -m\\
         m^* & 0 &  \Delta & 0\\
         0 & -m^* & 0 & -\Delta \\
         \end{array}
\right). \eea where \bea m &=& \frac{1}{4}(m_A+i m_B-m_C -im_D) \nn \\ &=& \frac{1}{4}(m_A+ e^{iQ} m_B +  e^{2iQ}m_C +  e^{3iQ} m_D) ,\eea and \bea \Delta= \frac{1}{4}(m_A+m_B+m_C+m_D). \eea More concisely, we have \bea \bar{H}'_{\rm eff} = m\sigma_z\tau_+ + m^*\sigma_z\tau_- +\Delta \sigma_z \eea
The full low-energy Hamiltonian is \bea \bar{H}_{\rm eff} &=& \bar{H}_{0,{\rm eff}} + \bar{H}'_{\rm eff} \nn \\ &=&  (v_x q_x \sigma_x + v_y q_y \sigma_y + v_z q_z \sigma_z \tau_z) \nn \\ && + ( m\sigma_z\tau_+ + m^*\sigma_z\tau_-) +\Delta \sigma_z. \eea All terms anticommute with each other except the last term $\Delta\sigma_z$. The physical effect of a nonzero $\Delta$ is to shift the location of the Weyl points, which does not significantly modify our results as long as $|\Delta|<|m| $ ( see the paper for discussions on the effects of $\Delta\sigma_z$, and a derivation of the condition $|\Delta|<|m| $).

As discussed in the paper, the Dirac mass for the four cases (I) $(m_A, m_B, m_C, m_D)=(m_1,m_2,m_3,m_4)$; (II) $(m_A, m_B, m_C, m_D)=(m_4,m_1,m_2,m_3 )$; (III) $(m_A, m_B, m_C, m_D)=(m_3,m_4,m_1,m_2)$; (IV) $(m_A, m_B, m_C, m_D)=( m_2,m_3,m_4,m_1)$, is given by $m_{\rm I}=(m_1+im_2   -m_3 -i m_4)/4$, $m_{\rm II}= (m_4+im_1   -m_2 -i m_3)/4=im_{\rm I}$, $m_{\rm III}= (m_3+im_4   -m_1 -i m_2)/4=-m_{\rm I}$, and $m_{\rm IV}= (m_2+im_3   -m_4 -i m_1)/4=-im_{\rm I}$, respectively. Translating the system by one site along the $z$ direction generates a phase factor $e^{iQ}=i$ for the Dirac mass, namely, $m\rw im$.

\begin{figure}
\centering
\includegraphics[width=9cm, height=3.8cm]{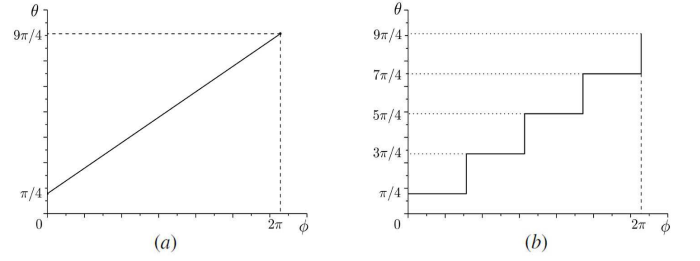}
\caption{   The evolution of the phase $\theta$ of the Dirac mass around the vortex line, for (a) an ideal vortex line of mass, and for (b) the discrete version of vortex line of mass in our paper. Here $\phi$ is the angular variable in the
 cylindrical coordinates $(z, r, \phi)$ ( $r=0$ is the location of vortex line ). The rough profiles of the evolution of $\theta$ are similar in (a) and (b), which justifies regarding the vortex line in our paper as a discrete analogue of the vortex string in continuous field theory.
  } \label{mass}
\end{figure}

In the paper, we have described the construction of a vortex line of mass.
For the case $(m_1,m_2,m_3,m_4)=(m_0,m_0,-m_0,-m_0)$, the evolution of the phase of the Dirac mass around the vortex line is shown in Fig.\ref{mass}b.

The phenomenon that a translation generates a phase for the Dirac mass has an analog in the Su-Schrieffer-Heeger(SSH) model (see Ref.\cite{fradkin2013} for an introduction). The SSH model describes a one-dimensional chain with hopping $t_{i,i+1}= t + (-1)^i m/2$ between the adjacent sites $i$ and $i+1$, in other words, for two neighboring links, the hopping amplitude is $t+m/2$ and $t-m/2$, respectively. The low energy theory of the SSH model is an (1+1)D Dirac fermion, the Dirac mass being equal to $m$. Translating the system by one lattice site generates a phase factor $\pi$ for the Dirac mass, namely, $m\rw -m$, therefore, there is no difference in the bulk properties to distinguish the $m>0$ and the $m<0$ configurations. Despite of this, if we create a domain wall between these two configurations, there will be zero modes living on it. In two-spatial dimensions, a related model on the honeycomb lattice has been studied\cite{hou2007}.

\section{Effect of incommensurate $k_0$ and $Q$}

Now we discuss the effect of shifting $k_0$ away from $\pi/4$, say $k_0=\pi/4+\delta$, $\delta$ being small. The basic idea is not difficult to explain. Suppose that $k_0=\pi/4+\delta$, $\delta$ being small, and the full Hamiltonian of this system is denoted as $H_\delta + H'$ (namely, $H_0$ is replaced by $H_\delta$). We split the Hamiltonian as \bea H_\delta + H' = H_0 + H' + (H_\delta - H_0), \eea and treat the $H_\delta - H_0$ term as perturbation. Let us first ignore the $H_\delta - H_0$ term, then the problem is just the case we have studied, and the explanation of the dislocation line as discrete vortex line is valid. As we have shown, there are unidirectional modes in this case. Then we take the $H_\delta - H_0$ term as a perturbation to this Hamiltonian. Since the Hamiltonian $H_0+H'$ has already an energy gap $\sim 2|m|$ in the bulk samples, the effects of perturbation is insignificant as long as it is not too big. In accordance with this insignificance of perturbation in the bulk sample, we do not expect the topologically robust unidirectional modes to suddenly disappear when the perturbation is small. Moreover, according to Callan and Harvey, the existence of unidirectional modes is consistent with the anomaly inflow in the bulk material; since an insignificant perturbation in the bulk material should not eliminate the anomaly inflow,  it should not eliminate the unidirectional modes either.

Now we give a more quantitative criterion.
In the enlarged unit cell, nonzero $\delta$ amounts to replacing $H_{ii}$ in $\bar{H}_0$ by \bea H_{ii}+[-2t_z\cos(\pi/4+\delta)+2t_z\cos(\pi/4)]\sigma_z,\eea in other words, \bea \bar{H}_0 \rightarrow
\bar{H}_0 +   \left( \begin{array}{cccc}
        F\sigma_z & 0 & 0 & 0 \\
         0 & F\sigma_z & 0 & 0\\
         0 & 0 & F \sigma_z & 0\\
         0 & 0 & 0 & F\sigma_z \\
         \end{array}
\right),  \eea where $F\equiv -2t_z\cos(\pi/4+\delta)+2t_z\cos(\pi/4)\approx  v_z\delta$. With this term included, we can repeat the previous derivation of $\bar{H}_{\rm eff}$ and obtain the new low-energy Dirac Hamiltonian for $k_0=\pi/4+\delta$:  \bea \bar{H}_{\rm eff}   =  (v_x q_x \sigma_x + v_y q_y \sigma_y + v_z q_z \sigma_z \tau_z) \nn \\ + ( m\sigma_z\tau_+ + m^*\sigma_z\tau_-) + (\Delta + v_z\delta) \sigma_z. \eea The energy eigenvalues are
\bea E_{1,2,3,4}(\bq)= \pm \sqrt{v_x^2 q_x^2 +v_y^2 q_y^2 + (\Delta+v_z\delta \pm\sqrt{v_z^2 q_z^2+|m|^2})^2},\eea which become gapless when $|\Delta + v_z\delta|\geq |m|$. As long as $|\Delta + v_z\delta| < |m|$, the bulk sample remains gapped, and the topologically robust unidirectional dislocation modes are expected to persist.

\section{Two-band low-energy model for the gyroid photonic Weyl crystals}

In the paper we have studied a simplified two-band model and show that vortex lines of Dirac mass carry unidirectional modes. Since our central proposal is to realize unidirectional modes in photonic Weyl materials, we would like to show that this model is indeed relevant to the experimentally feasible materials, the gyroid photonic crystals.

Gyroid photonic Weyl materials have been proposed in Ref.\cite{lu2013weyl} and observed in Ref.\cite{lu2015}. The low energy physics of these photonic Weyl materials can be described by a three-band model (see the supplemental material of Ref.\cite{lu2013weyl}).
Here, we would like to reduce the three-band model to a two-band model resembling the one studied in our paper.

In the presence of a magnetic field $\bB$, the low-energy Hamiltonian is given as\cite{lu2013weyl} \bea H(\bk)=\alpha_1|\bk|^2 -\alpha_2(\bk\cdot\bL)^2 +\beta(\bB\cdot\bL), \eea where  $\bL=(L_x, L_y, L_z)$ with \bea L_x= -i
             \left(
                \begin{array}{ccc}
                  0 & 0 & 0   \\
                   0 & 0 & 1 \\
                   0 & -1 & 0  \\
                \end{array}
 \right) \nn \\
 L_y= -i   \left(
                \begin{array}{ccc}
                  0 & 0 & -1   \\
                   0 & 0 & 0 \\
                   1 & 0 & 0  \\
                \end{array}
 \right) \nn \\
 L_z= -i\left(
                \begin{array}{ccc}
                  0 & 1 & 0   \\
                   -1 & 0 & 0 \\
                   0 & 0 & 0  \\
                \end{array}
 \right) \nn
\eea
This low-energy Hamiltonian harbors a pair of Weyl points at $\bk=(0,0,\pm k_0)$, as we now show. Hereafter, we take $\bB=(0,0,B)$ with $\beta B>0$.  By an explicit calculation, we have \bea  H((0,0,k_z)) =  \left(
                \begin{array}{ccc}
                  (\alpha_1-\alpha_2)k_z^2 & -i\beta B & 0   \\
                   i\beta B & (\alpha_1-\alpha_2)k_z^2 & 0 \\
                   0 & 0 & \alpha_1 k_z^2  \\
                \end{array}
 \right),
 \eea along the line $\bk=(0,0,k_z)$.
The three eigenvectors are $|1\ra = (0,0,1)^T$, $|2\ra=(1,i,0)^T/\sqrt{2}$,
and $|3\ra = (1,-i,0)^T/\sqrt{2}$, with energies \bea E_1 &=&\alpha_1 k_z^2,
\nn \\ E_2 &=& (\alpha_1-\alpha_2)k_z^2 + \beta B, \nn \\ E_3 &=&
(\alpha_1-\alpha_2)k_z^2 -  \beta B. \eea When $k_z=\pm k_0 \equiv\pm
\sqrt{ \beta B/\alpha_2}$, we have the band touching $E_1=E_2$, which
defines the Weyl points. By tuning the parameters of the gyroid photonic Weyl crystals, one can tune $\bk_0$ to $(0,0,\pi/4)$.  Since $E_3$ is far away from $E_1, E_2$, we project the Hamiltonian $H(\bk)$ to the subspace spanned
by $|1\ra$ and $|2\ra$. Let us introduce three Pauli matrices $\sigma_{x,y,z}$ in this two-dimensional space.  Near $\bK_\pm = (0, 0, \pm k_0)$, we find that the effective two-band Hamiltonian reads \bea H_{{\rm eff}} = \pm k_0[\frac{1}{\sqrt{2}} \alpha_2  (q_x\sigma_x -q_y\sigma_y) -\alpha_2  q_z \sigma_z +(2\alpha_1-\alpha_2) q_z] \nn \\ + \cdots, \eea where $q_i\equiv k_i - (\bK_\pm)_i$, and ``$\cdots$'' denotes higher order terms.
Omitting the last term, which is proportional to the unity matrix, we have \bea  H_W = \tau_z\sum_i v_i\sigma_i q_i, \label{photonic-2-band} \eea where $\tau_z=\pm 1$ refers to $\bK_\pm$, $v_x=-v_y= \frac{1}{\sqrt{2}} \alpha_2 k_0$, and $v_z= -\alpha_2 k_0$.   Here,  Eq.(\ref{photonic-2-band}) takes almost the same form as Eq.(2) in the paper,  except for the inessential difference that $\tau_z$ appears in all the three terms (instead of appearing only in one of these three terms). The Dirac mass term for Eq.(\ref{photonic-2-band}) is \bea H_M =m\tau_+ + m^*\tau_-, \eea where $\tau_\pm =(\tau_x\pm i\tau_y)/2$. Compared to the $H_M$ given in the paper, there is no $\sigma_z$ factor here. This is due to the slight difference in the form of $H_W$. In each case, the Dirac mass term anticommutes with all terms in $H_W$. To generate such a Dirac mass, we only need to add the following term: \bea H'=\sum_j m_j\sigma_0(j), \eea where $\sigma_0$ is the $2\times 2$ unit matrix. Repeating the calculation in Sec.\ref{calculation} of this Supplemental Material, we can obtain the Dirac mass \bea m &=& \frac{1}{4}(m_A+i m_B-m_C -im_D) \nn \\ &=& \frac{1}{4}(m_A+ e^{iQ} m_B +  e^{2iQ}m_C +  e^{3iQ} m_D). \eea The simplest choice is $(m_A, m_B, m_C, m_D)=(m_0,m_0,-m_0,-m_0)$, which leads to $m=m_0\exp(i\pi/4)/\sqrt{2}$.

In the low-energy theory, the presence and absence of each term can be determined by symmetries.
From symmetry consideration, any modulation of the photonic lattice with the $A,B$ unit cells different from the $C,D$ unit cells will produce the $H'$ term in the low-energy effective theory, thus it will generate a Dirac mass, though the precise form of the band structure and the precise value of $m$ can be determined only by photonic band structure calculations, which is beyond the scope of the present work. The existence of unidirectional modes along the vortex line of mass is nevertheless a topologically robust phenomenon.

In conclusion,  in this last section of the Supplemental Material, we have shown that the two-band model we have studied in the paper (and its slight modification) captures the low-energy physics of the gyroid photonic Weyl materials, therefore, our prediction of unidirectional modes along the vortex line of mass should be observable in these materials.

\end{document}